**Programmed Internal Reconfigurations in a 3D-Printed Auxetic Metamaterial Enable Fluidic Control for a Vertically Stacked Valve Array**

**AUTHORS**

Tinku Supakar,[1] David Space,[1] Sophy Meija,[2] Rou Yu Tan,[3] Jeffrey R. Alston,[4] and Eric A. Josephs[1,2,*]

[1]Department of Nanoscience, Joint School of Nanoscience and Nanoengineering, University of North Carolina at Greensboro, Greensboro, NC, USA 27401

[2]Department of Biology, College of Arts and Sciences, University of North Carolina at Greensboro, Greensboro, NC, USA 27412

[3]The Early College at Guilford, Greensboro, NC, USA 27410

[4]Department of Nanoengineering, Joint School of Nanoscience and Nanoengineering, North Carolina Agricultural and Technical State University, Greensboro, NC, USA 27401

* To whom correspondence should be addressed. Tel: +1 336 285 2890; Email: eric.josephs@uncg.edu

**ABSTRACT**

Microfluidic valves play a key role within microfluidic systems by regulating fluid flow through distinct microchannels, enabling many advanced applications in medical diagnostics, lab-on-chips, and laboratory automation. While microfluidic systems are often limited to planar structures, 3D printing enables new capabilities to generate complex designs for fluidic circuits with higher densities and integrated components. However, the control of fluids within 3D structures presents several difficulties, making it challenging to scale effectively and many fluidic devices are still often restricted to quasi-planar structures. Incorporating mechanical metamaterials that exhibit spatially adjustable mechanical properties into microfluidic systems provides an opportunity to address these challenges. Here, we have performed systematic computational and experimental characterization of a modified auxetic structure to generate a modular metamaterial for an active device that allows us to directly regulate flow through integrated, multiplexed fluidic channels "one-at-a-time," in a manner that is highly scalable. We present a design algorithm so that this architecture can be extended to arbitrary geometries, and we expect that by incorporation of mechanical metamaterial designs into 3D printed fluidic systems, which themselves are readily expandible to any complex geometries, will enable new biotechnological and biomedical applications of 3D printed devices.  (194 words)

**INTRODUCTION**

Microfluidic valves play a key role as integrated elements within microfluidic circuits,[1-2] regulating fluid flow in microchannels and enabling a diverse array of applications across fields such as analytical chemistry,[3-4] biotechnology,[5-7] and in medical diagnostics.[7-8] Many of these sectors, especially in drug discovery[9-10] and high-throughput molecular cell biology,[11-12] benefit significantly when equipped with valving systems that can be effortlessly expanded to accommodate a vast array of multiplexed fluidic inputs and outputs.[13-14] To achieve this type of scalable control across a considerable number of fluidic valves, a common approach relies on an architecture[15-16] whereby one valve can be designed to shut off flow across multiple channels simultaneously (Figure 1A-C). In a design pioneered by Quake *et al*.[17-18] the channels carrying the fluid of interest (the "flow" channels) are arranged in parallel beneath pneumatically-actuated "control" channels. These control channels are designed to close specific flow channels at regions of overlap ("closeable" segments) when pressure is increased. Simultaneously, other flow channels remain open regardless of the pressure in the control channel at the regions of overlap ("always open" segments). This configuration can be designed so that only a single flow channel is open when any combination of half of the control channels is pressurized (Figure 1C). Consequently, it enables the addressing of $N!/(N/2)!^2$ individual flow channels using only $N$ control channels. For instance, with 6 control channels, it can regulate 20 flow channels, so only one is open at a time; with 8 control channels, it can regulate 70 individually-addressable flow channels; and with 10 control channels, it can regulate 252 individually-addressable flow channels, and so forth, scaling factorially.

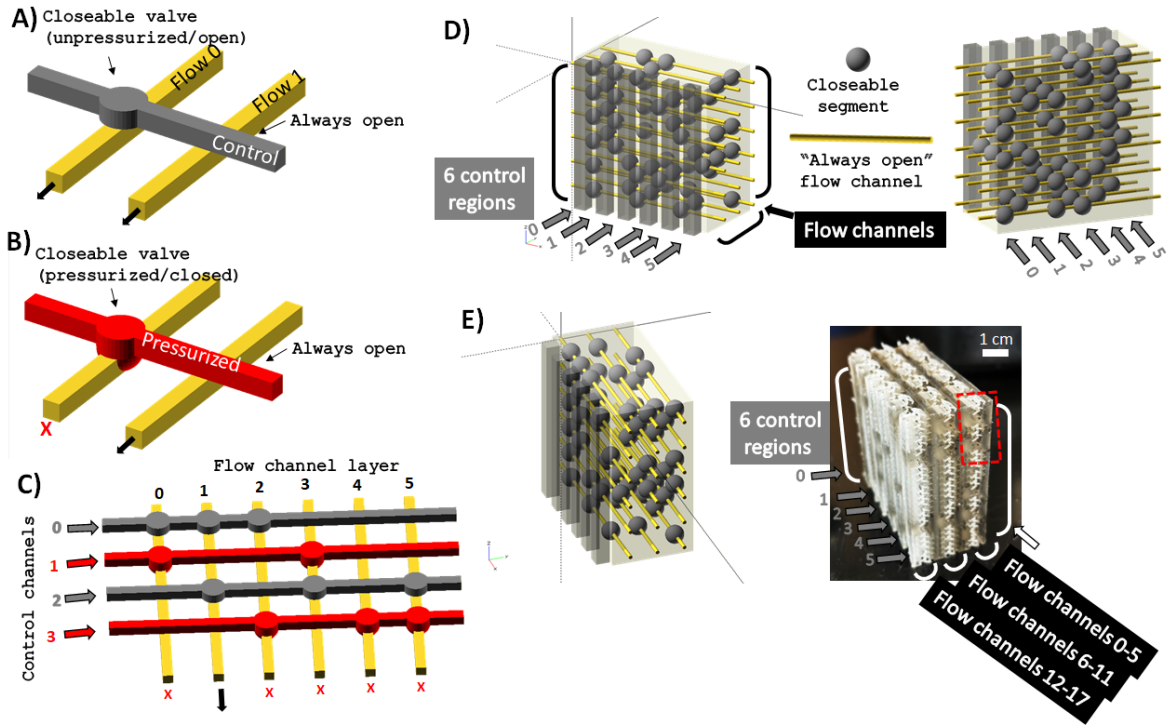

*Figure 1. A 3D-printed, vertically stacked valve array for fluidic control.* A+B) In traditional, quasi-planar fluidic valves arrays (with "Quake-style" valves), pneumatic "control" channels are placed above a series of "flow" channels. When specific control channels are pressurized (red), "closeable" segments of the flow channels that lie underneath the pressurized control channels collapse to obstruct flow, while other flow channels ("always open") underneath the same control channel can remain unobstructed regardless of its state (pressurized/unpressured. C) This architecture allows for N control channels to regulate $N!/(N/2)!^2$ for any combination of N/2 pressurized control channels. Here, closeable valves are arranged so that pressurizing any 2 of 4 control channels allows flow through only 1 of 6 flow channels by blocking flow through the other 5. D) Schematic (and reverse, right) of a stacked valve array in 3 dimensions, where collapsible regions of flow channels are positioned so that applying pressure to any 3 of 6 "control regions" allows flow through only 1 of 18 flow channels in the device. Vertical bars highlight the planes of the "control regions" in the device. E) Schematic and 3D printed valve array manifold, engineered using a mechanical metamaterial design to control the propagation of forces within its structure for the controllable collapse of integrated flow channels.

Fabricating these microfluidic systems typically requires a complex, multilayer process to properly arrange and register control channels atop the flow channels, separated by a thin membrane;[13, 16-20]; however, advances in additive manufacturing with 3D printing have opened new opportunities to make these devices outside of a clean room facilities. 3D printing has enabled the rapid fabrication and prototyping of miniaturized, intricately designed components which can control the flow of fluids at the microscale level,[21-34] and allows for the creation of monolithic, integrated microfluidic devices without any leakage issues.[23, 26-27] Digital light processing (DLP) 3D printing,[35-36] which builds objects layer by layer using a planar light source, is highly versatile for scaling as it allows simultaneous printing of large number of valves and channels. Even "Quake-style" valves have been generated using DLP 3D printing,[26] albeit using a custom-made high-resolution 3D printer and specialized 3D printing resins.

Despite these advances, most 3D printed microfluidic systems remain essentially quasi-planar structures,[37] meaning they do not fully take advantage of the scaling capabilities offered by 3D printing. This limitation can likely be attributed to the complexity associated with regulating pressure propagation in three dimensions to control valve actuation within the device.[15] As a result, these quasi-planar devices are limited in their ability to achieve higher throughput, as the channel dimensions of these 3D printed valves are restricted by the resolution of commercially available 3D printers. Here we have aimed to design a 3D-printed valve array system that are stacked in 3 dimensions and that can be easily scaled, like "Quake-style" valving" system,[17-18, 38] using auxetic mechanical metamaterial structures (Figures 1C-D).[39-41] After systematic experimental and computational characterization of auxetic structures, we have found that through the adjustment of a single characteristic of its auxetic cells, its 'tilt' angle θ (Figure 2A-C), we can generate metamaterial structures with specific regions that either remain stable or collapse under applied forces . This capability allows us to generate 3-dimensional structures with engineered "always open" and "closeable" valve regions (Figure 1C) that can be stacked to build arrays of "Quake"-like valves within a 3D printed framework. We also present an algorithm to design and print these "stacked" fluidic arrays with arbitrary geometries, and to illustrate this, we present a fluidic array of 6 x 3 (18 total) flow channels that are individually controllable through the application of forces on any 3 out of 6 "control" regions on the structure (Figure 1D), and demonstrate how these devices can be expanded even further (*e.g.* to 96 flow channels regulated by 9 control regions). While auxetic structures are often used as "passive" materials with robust stretching and tunable deformation characteristics, we expect that the integration of active and dynamic auxetic mechanical metamaterials with microfluidic devices that is enabled by advances in the 3D printing will allow for new applications in lab-on-chips and biotechnology.[21, 29]

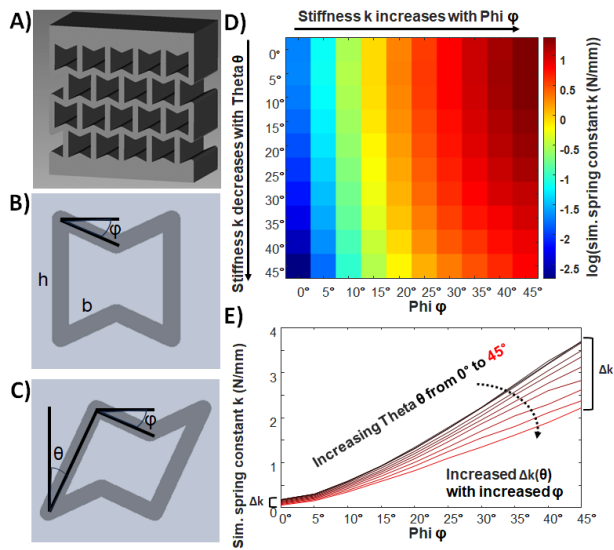

*Figure 2. The "tilt" angle in a re-entrant auxetic metamaterial provides a handle to tune relative mechanical stiffness within a 3D printed device. A) The structure of a re-entrant mechanical metamaterial, with a "bowtie" structure offset by half the cell width in each layer. B) The basic unit of a re-entrant auxetic, defined by the re-entrant angle phi φ which defines its key mechanical property of a negative Poisson ratio (NPR). h is the length of the side wall and b is the length of the re-entrant arm. C) We hypothesized that tilting the unit cells with tilt angle theta θ allows successive layers with the same φ to stack and maintain registry while providing an additional handle on the mechanical stiffness of the layer. D) According to finite element analysis (FEA) of re-entrant auxetic unit cells subjected to forces, the estimated (simulated) stiffness increases with increased φ but decreases with the same φ with increased θ (larger tilt). E) Increasing φ allows for greater differences in stiffness between layers with different tilts, a prerequisite for engineering "always open" and "closeable" segments of a fluidic device.*

## RESULTS

In expanding a "Quake-style" valve array into the third dimension, we require the ability to generate "closeable" segments and "always open" segments of flow channels through the device to which a pressure can be applied in parallel across the entire device (Figure 1D). This required the design of a material with controllable force propagation throughout the device so that all valves within a "control" region could be addressed simultaneously through the 3D structure. Initial designs focused using auxetic re-entrant metamaterials (Figures 2A-B), which are mechanical metamaterials with a repeated cellular structure in the shape of overlapping bowties, as they are commonly used in applications where robust controllable deformation is required.[40-41] Auxetic metamaterials have the unusual property of negative Poisson ratios determined by their "reentrant angle" $\varphi$.

Finite element analysis (FEA) predicted that the stiffness of the auxetic materials would increase with increasing $\varphi$ (Figures 2D-E). This finding was consistent with initial experiments to test force *vs.* displacement using 3D printed 5x5 auxetic grids (Figures 3A-B and S1; which are 4 layers of 5 auxetic cells capped by 2 half-layers of solid, "filled-in" auxetic cells). We hypothesized that by changing the tilt ($\theta$) of the reentrant structure (Figure 2C), we could modulate mechanical properties of the metamaterial in such a way that the auxetic structures, regardless of their stiffness, would still maintain registry across layers by virtue of their identical $\varphi$ angles (i.*e.*, Figure 3C). The results of FEA of individual auxetic cells predicted that increasing $\theta$ would result in more pliant structures (lower simulated spring constants $k$), as expected. Further, it revealed that with increased $\varphi$, the differences in spring constants ($\Delta k$) between auxetic cells with different $\theta$ also increased (Figures 2D-E). Building on this principle, FEA revealed that sandwiching 2 tilted layers between un-tilted layers ($\theta = 0°$) would result in greater deflections for the same force (Figure 3C). Further, compared to un-tilted structures—where initial stresses were evenly spread through the structure, and where, regardless of $\varphi$, all layers collapse simultaneously (Figures 3B and S1)—the stresses would be directed through the tilted layers upon the initial application of force (Figure 3C). Defining regions of preferential compliance or rigidity in three dimensions forms the basis of how we will generate "closeable" or "always open" segments for integrated flow channels.

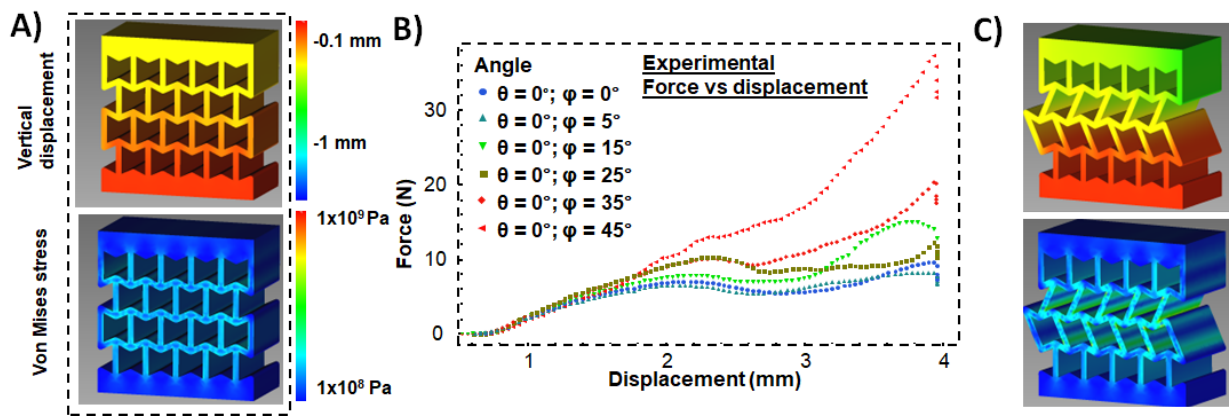

***Figure 3. Tilting re-entrant auxetic metamaterials focuses mechanical stresses within the structures.*** *A) FEA of 5x5 un-tilted ($\theta = 0°$, $\varphi = 25°$) auxetic grids reveals that stresses are distributed evenly throughout. B) Experimental force vs. displacement curves of un-tilted 5x5 auxetic grids (as in 3A, dotted box) with different $\varphi$ reveals that all layers deform simultaneously upon compression (see Figure S1) and that these structures are significantly stiffer than those with tilted interior layers (see Figure 4). C) FEA reveals that introducing tilted auxetic cells ($\theta = 25°$, $\varphi = 25°$) between un- tilted ($\theta = 0°$, $\varphi = 25°$) layers focuses the stress to the tilted layers initially and results in greater initial deflections for the same force. Coloration the same as in 3A.*

To maximize the difference in mechanical responses between un-tilted (Figure 4A) and tilted (Figure 4B) layers, we made two adjustments: 1) we "hollowed" the tilted auxetic cells, giving them a wire-frame structure in three dimensions, further increasing their compliance, while un-tilted cells had solid walls, and 2) we systemically varied θ (for tilted cells) and φ (for both tilted and un-tilted cells, to maintain registration) to optimize these "banded" structures for use as "closeable" or "always open" segments in a 3D printed valve array (Figures 4C, S2, and S3). The 5x5 "banded" structures are built with (Figures 3C and 4D):

1) Half of one a layer of a solid, filled-in untiled auxetic cells on the top and half of a layer of solid, filled-in untiled auxetic cells on the bottom of the structure;
2) a layer of untilted, solid-walled untilted auxetic cells with empty interiors;
3) one layer of tilted, wire-framed auxetic cells;
4) one layer of wire-framed auxetic cells tilted at the same angle in the opposite direction;
5) one layer of untiled solid walled auxetic cells with empty interiors

Note that as part of our design criteria, we want each 5x5 grid to remain the same width ($W$) and length ($L$) (here, 9 mm x 9 mm) regardless of θ (for tilted cells) and φ. To do so, the side length of the bowties ($h$ in Figure 2b) and re-entrant arm length ($b$) were adjusted for each set of conditions so that $b = W / (8*\cos(\varphi))$ and $h = (L + 5*b*\sin(\varphi)) / (3+2*\cos(\theta))$, allowing the footprint of each 5x5 structure remains the same overall.

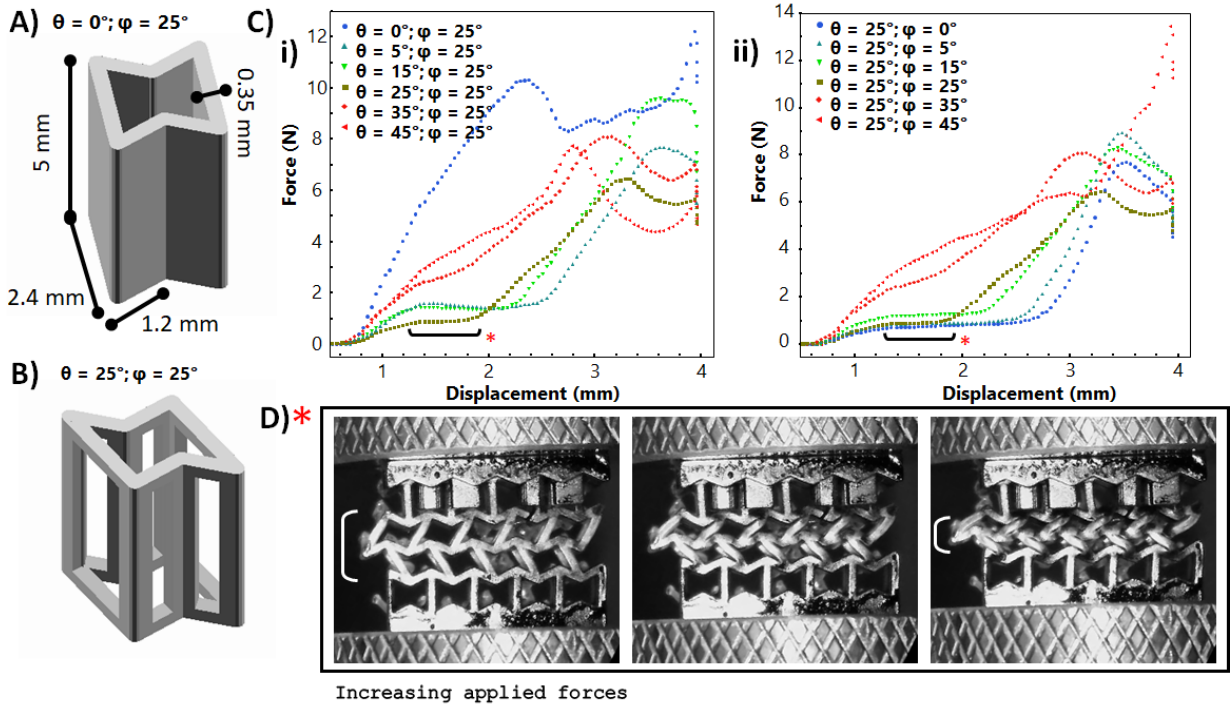

*Figure 4. Banded (tilted) and hollowed re-entrant auxetic metamaterials allow for controllable collapse of the specific regions in the 3D printed structures. A) 3D ('extruded') structure of the un-tilted re-entrant layers. B) Tilted layers are given 'wireframe' structures to further enhance their compliance compared to the un-tilted layers. C) Experimental force vs. displacement curves for banded re-entrant 5x5 grids auxetic with un-tilted outer layers and tilted interior layers (see Figure 2Fii, for example), for increasing tilt angle theta θ for a given re-entrant angle phi φ (i) or increasing φ for a given θ. D) The interior layers (θ = 25°, φ = 25°) of banded tilted re-entrant auxetic metamaterials completely collapse before the outer layers (θ = 0°, φ = 25°) buckle, while maintaining spatial registration in the material.*

We found the inner, tilted layers collapsed more readily as θ increased from 0° to 30°, after which the structures became more rigid and the inner and outer layers collapsed together (Figures

4Ci and S2). Structures with θ = 25° have a minimum in the force required to collapse the interior layer, as those with higher θ tend to jam quickly; structures with lower φ require more displacement in order for the middle layers to completely collapse, while those with higher φ are significantly stiffer and cause the outer walls to buckle during their collapse. Conversely, for fixed θ values, the auxetic structures exhibit the highest compressibility at the lowest φ value (zero degrees) and become less compressible as the value of φ increases (Figures 4Cii and S3).

Structures with tilted layers having θ = 25°, φ = 25° (brown squares in Figure 4C) provides a 'sweet spot', with a short, low-force plateau during which the interior collapses first before the outer layers begin to deform (Figure 4D). Additionally, when assembled into larger arrays, structures with higher θ values (>25° degrees) tended to buckle irregularly when subjected to applied forces (Figure S4). Therefore, auxetic grids with bands of tilted layers having θ = 25°, φ = 25° formed the basis of the "closeable" and "always open" segments of integrated flow cells (Figure 5A). To engineer closeable segments, we positioned solid segments around integrated flow channels within the auxetic grids, such that applied forces would call the interior region to collapse and seal the flow channel at that segment. The "always open" segments also collapsed, but redirected forces around the flow channel so it would remain open. In this way, if we apply mechanical pressures across stacked arrays of flow channels in parallel (at a "control region", Figure 1D), depending on the structure around the flow channel it will either collapse or remain open (Figure 5B). When pressure was applied simultaneously across both segments, the flow channels in the closeable segments collapsed and flow of liquids through that channel ceased, while flow through the always-open segments was unaffected (Figure 5C). Multiple "closeable" and "always open" segments could be operated in parallel and independently when pressure is applied across a common control segment (Figure 6A). As an initial demonstration of controlled valving, we then constructed a 3D printed quasi-planar "Quake-style" manifold (as in Figure 1C) using the auxetic metamaterial segments with 4 control regions governing 6 channels, where pressure on any 2 control regions only allowed flow in a single region (Figures). This manifold could be integrated to control inputs into a microfluidic chip or into fluidic multiplexors / de-multiplexors (Figures 6B-D and S5). Having developed "always open" and "closeable" segments of flow channels, these features allow for the controlled propagation of mechanical forces through a device and can be stacked to create a 3D, multi-layer valve array (Figure 7).

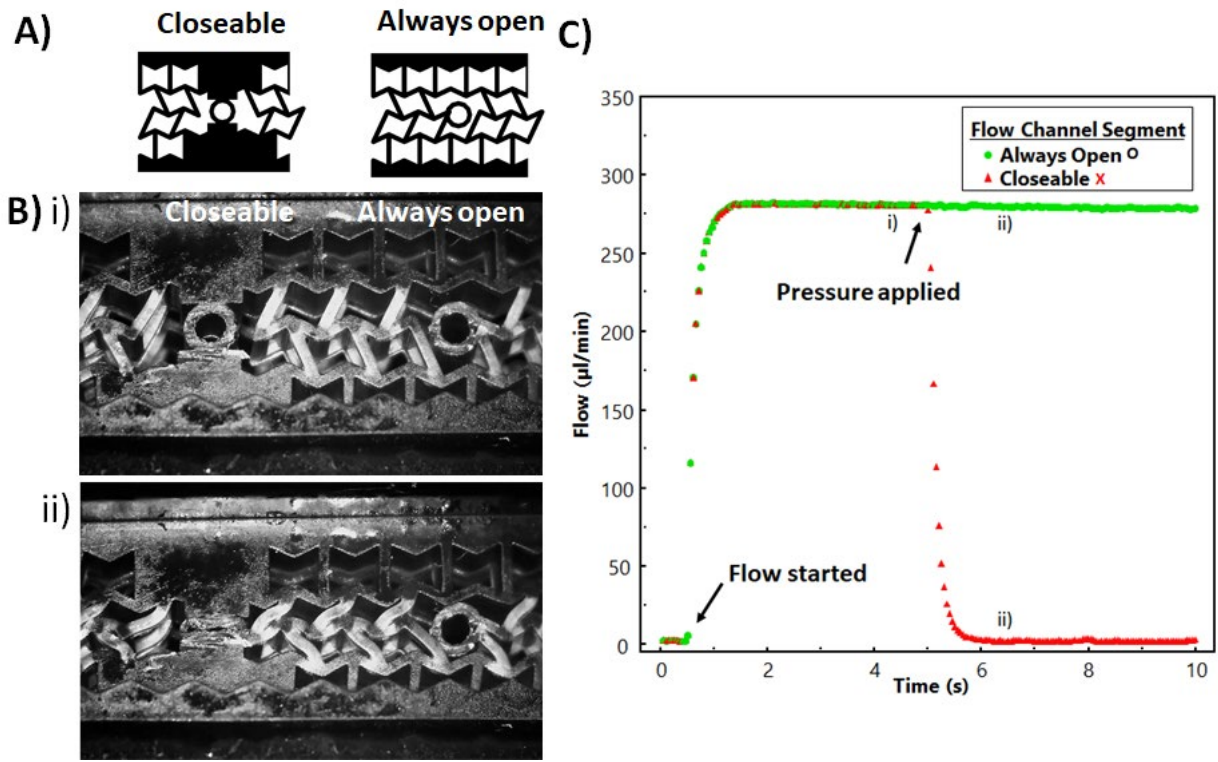

*Figure 5. Tilted re-entrant auxetic mechanical metamaterials allow for the generation of "always open" and "closeable" segments / valves when subjected to applied pressures in parallel. A) 2D-projection of the design of "closeable" and "always open" flow channel segments. The flow channel is the made from the circular segment in the middle of each design. Note that the un-tiled and (hollowed/wire-framed) tiled auxetic cells have a 3D structure as shown in Figures 4A-B. B) (above) Before pressure is applied, both fluidic channels channels are open. (below) After pressure is applied to a "control region" above both segments, only the closeable channel collapses. C) Flow through two integrated channels (i) is high until both are simultaneously subjected to pressure applied across a "closeable" (red) and "always open" (green) segments until flow is obstructed in the flow channel with the closeable segment while flow in the channel with the always-open segment remains unchanged (ii).*

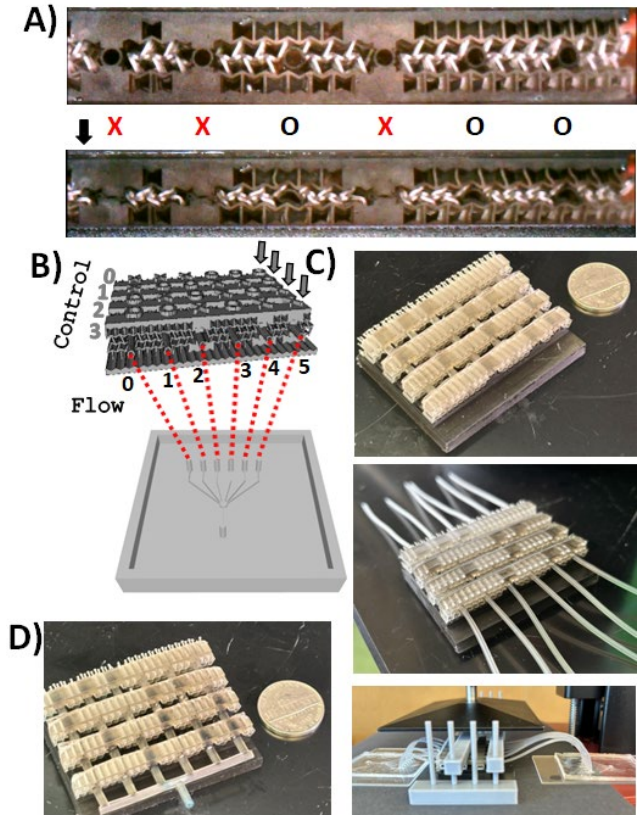

*Figure 6. A "quasi-planar" valve array integrates into microfluidic multiplexors / de-multiplexors. A) An arrangement of 6 integrated flow channels, 3 of which have always open ('O') segments at the control region and 3 have closeable ('X') segments at the control region. When subjected to pressure across the entire control region, (below) only the flow channels with closeable segments collapse and are completely obstructed. B) Design of a 3D printed (quasi-planar) 6 flow channel / 4 control region metamaterial manifold (with architecture similar as in Figures 1C) for integration with a microfluidic multiplexor. C-D) Practical integration of the 3D printed quasi-planar valve array into microfluidic multiplexor / de-multiplexor architectures. See Figure S5.*

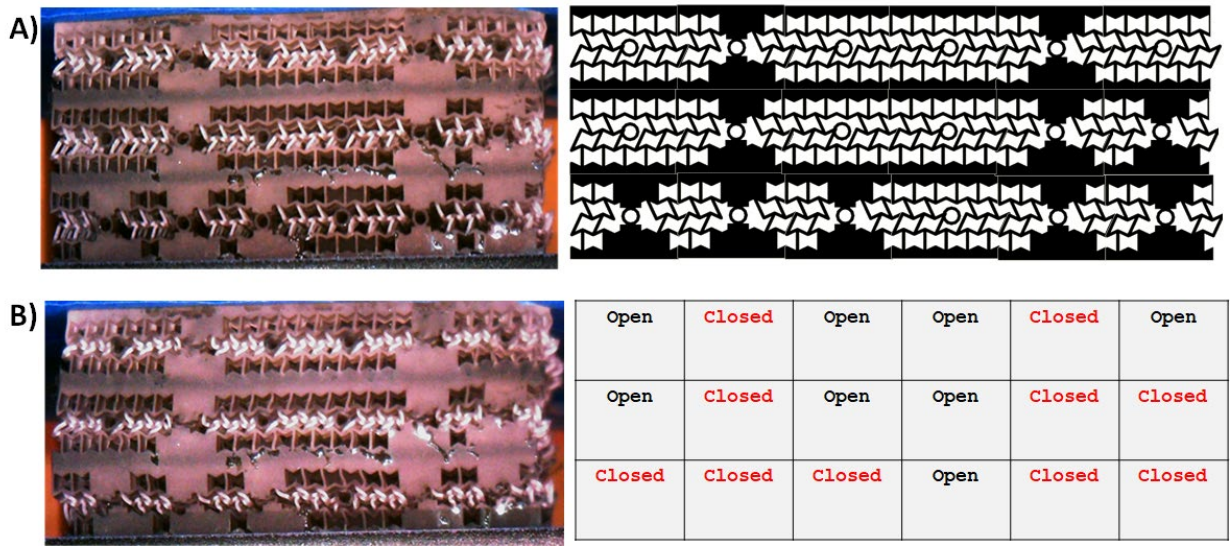

*Figure 7. A stacked valve array with complex positioning of "always open" and "closeable" segments of flow channels. A) A 3D printed stacked valve array (as in Figure 8E) has "always open" and "closeable" segments of 18 integrated flow channels distributed throughout, with some atop or between other kinds. (right) schematic projection of*

*a multi-layer "control region". Note that the "always open" and "closeable" segments (from Figure 5A) of the flow channel are modular and fit together. B) Upon application of pressure to a control region (from the top) the closeable segments all collapse throughout the device, regardless of its position or location relative to "always open" valves, which remain unobstructed. (right) State of the integrated channels after application of pressure.*

To design a multi-layer valve array of arbitrary geometry, the number of control regions $N$ for $X$ number of flow channels must satisfy $N!/(N/2)!^2 \geq X$, which we demonstrate here with a 18, individually-addressable (only 1 is open at a time) flow channel device operated by 6 control regions (6x3x6, Figure 8). For a set of flow channels, it is useful to design a scheme where the locations of "always open" and "closeable" can be algorithmically positioned at their precise control regions to allow individual addressability, and, if we wish specifically open a specific control channel, to be able to readily calculate the control regions upon which pressure should be applied. To do so, we enumerate all combinations of $N$ control channels (indexed from $0$ to $N - 1$) of length $N/2$ as $[n_1, n_2, \ldots n_k]$ (Figure 8A) for $k = N/2$ using the nested loop:[42-43]

```
for n₁ = 0 to N - k + 1
    for n₂ = (n₁ + 1) to N - k
        ...
            for nₖ = (nₖ₋₁ + 1) to N
                print [n₁, n₂,…nₖ]
```

that are each associated with a given flow channel with index $i = 0, 1, \ldots, X-1$ (the indexes of specific flow channels are enumerated starting at 0). For each flow channel, the "always open" segments are placed within the control regions whose indices are found within their corresponding combinations, while the closeable segments are placed within the control regions whose indices are not found in their combination (Figure 8A). This results in every flow channel having a unique combination of N/2 control regions that, when pressed, will result in every other flow channel becoming obstructed while that flow channel itself remains unobstructed. Having determined the positions for each flow channel in space (Figures 8B-C), the modular designs for the "always open" and "closeable" segments of the flow channels can be placed according to Figure 8A at within respective control regions/layers (Figure 8D). Because the "always open" and "closeable" segments are modular and fit with each other, once their geometry has been established, the segments can be positioned algorithmically according to this protocol using programmatic computer-aided design software in order to generate these complex, aperiodic/irregular three-dimensional structures for valve arrays or arbitrary designs or complexity that would be extremely difficult and cumbersome to design by hand.

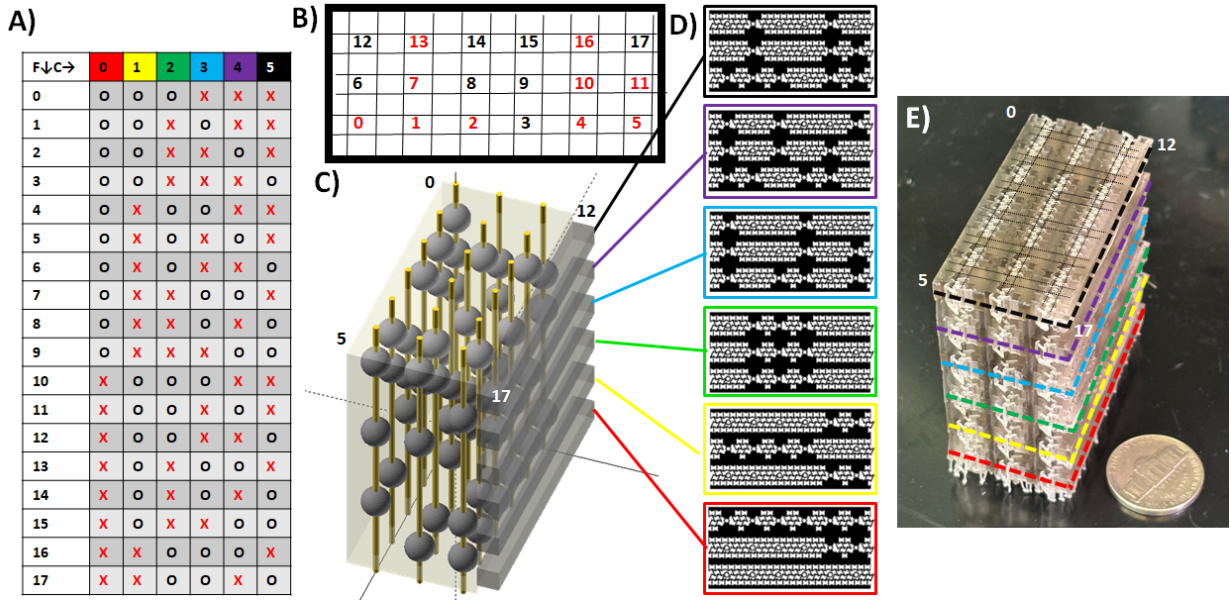

***Figure 8. Design of a 6x3x6 3D-printed, vertically stacked valve array with 18 independently addressable flow channels.*** *A) For an 18 fluid channel fluidic device (labeled from 0 to 17), we require 6 control regions (labeled from 0 to 5). To determine the positions of "closeable" and "always open" segments to uniquely allow only a single channel to be opened when pressure is applied to any 3 (= 6/2) control regions, first we enumerate all unique combinations of 3 control regions for, marking an "O" at those regions and an "X" for "closeable" segments at the regions not in that last. B) For 6x3 flow channels, the positions of each channel through a control region are mapped, for example, for control region 5 (black). C) The positions of closed channels (spheres, as in Figure 1D) and mapped to a three dimensional structure with control regions highlighted as dark gray bars. D) The 3D geometries for the modular "always open" and "closeable" segments (as in Figure 5A) are positioned within each independent layer. E) The design can then be 3D printed.*

In order to open a specific flow channel *i*, the combination of control regions to be compressed can be found using a look-up table (*i.e.* Figure 8A) so those respective control regions can be compressed. In the case of arbitrarily complex structures, the control regions to compress to open specific flow channel *i* can be quickly determined using the algorithm:[42-43]

```
1) Find max( c_k such that (c_k choose k) ≤ (N − 1 − i) )

2) Find max( c_{k-1} such that (c_{k-1} choose (k − 1)) ≤ (N − 1 − i) − (c_k
   choose k) )

   ...

3) Find max( c_1 such that (c_1 choose 1) ≤ (N − 1 − i) − (sum [(c_{k-j}
   choose k − j)] for j from [1 to k] ) )

4) Compress control regions [(N − 1) − c_1, (N − 1) − c_2,... (N − 1)
   − c_k]
```

where *(c choose k) = c!/((k!)*(c-k)!)*. The ability to algorithmically position different segments to generate individually-addressable arrays of flow channels and to algorithmically determine the inputs required to control any individual channel means this design approach can be used for arbitrarily complex structures, for example, a 96-flow channel array (9x12) requiring 9 control valves, such that any 4 of 9 will allow only a 1 of 96 channels to remain open (Figure 9).

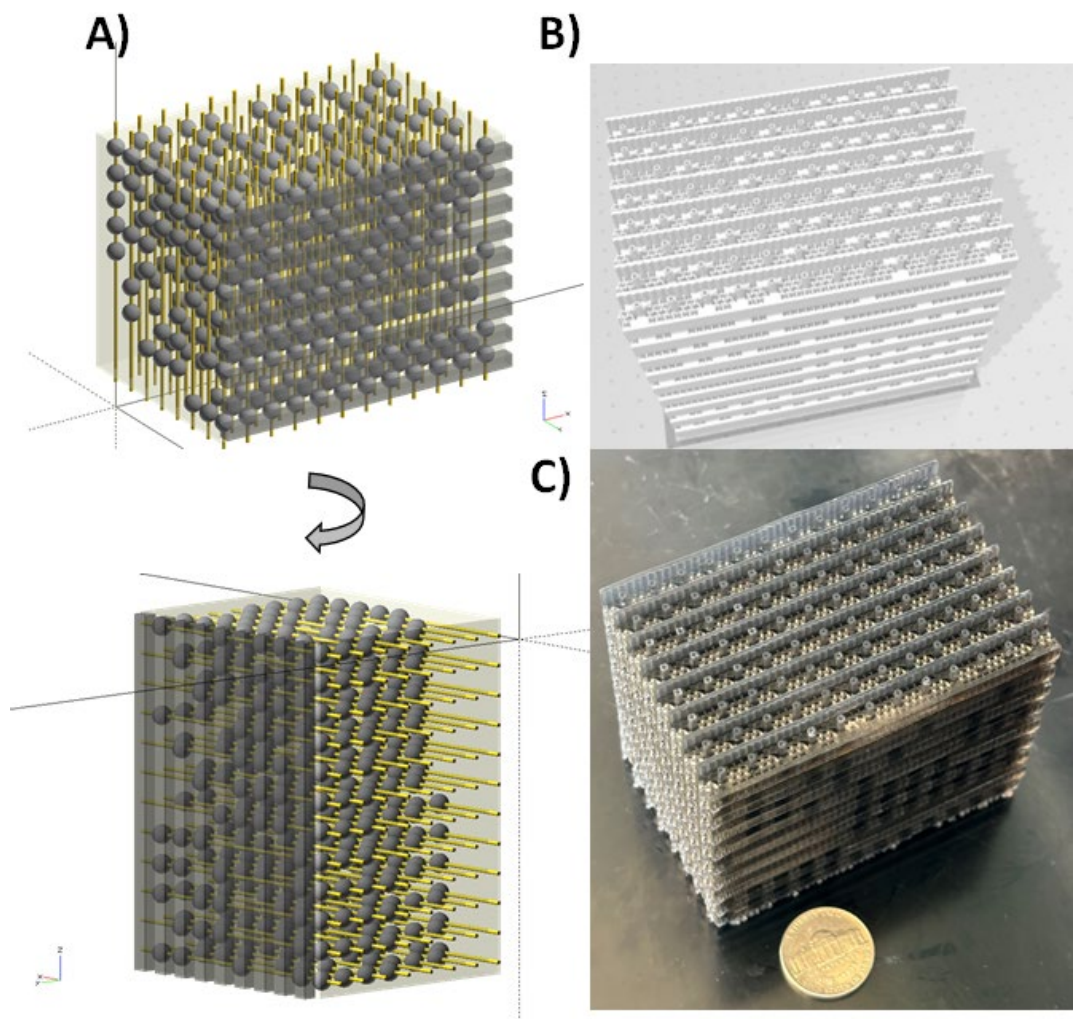

*Figure 9. Design and 3D-printing of a 9x12x9 vertically stacked valve array with 96 independently addressable flow channels.* Following the algorithm described in the text and demonstrated in Figure 8, multi-layer 3D printed valve arrays and nearly any geometry or complexity, such as this 96-flow channel valve array, can be algorithmically designed (A) and an irregular/aperiodic structure generated in three dimensions to position the "always open" and "closeable" segments within each control region (B). C) These intriquite structures can then be 3D printed using a DLP printer. The design can be made to fit above a standard 96-well plate.

Note that each control region contains a unique configuration of "always open" and "closeable" segments of each of the flow cells running through them, so that as complexity increases, algorithmic design of these modular segments becomes increasingly necessary.

## DISCUSSION

While auxetic mechanical metamaterials are often used "passively", where their unusual deformation properties are exploited for example to fit curved surfaces or minimize impact damage,[39-40] we show that integrating 3D printed, flexible mechanical metamaterials into complex, algorithmically-designed fluidic valve architectures provides a way to expand a "Quake-style" valve arrays and active fluidic control into the "third dimension". Doing so required precision positioning of solid and wire-framed (un-)tilted auxetic cells at specific 3D locations in a manner that allowed control of the propagation of applied forces and programmable collapse or opening of specific flow channels uniformly along the structure of the device. To our knowledge, "multi-layer Quake-style" architectures have not been demonstrated otherwise, but without 3D printing doing so would require extensive expertise, effort, and expense to generate in a clean-room environment. Our use of commercially-available, inexpensive DLP 3D printers and resins along with algorithmic design principles means that these technologies should be readily accessible, deployable, and customizable for different applications requiring fluidic control at various levels of complexity.

In principle, applying pressure to the control regions of these devices could be performed using automated mechanical actuators or pneumatic pressure, using open-source control systems that have already been designed for "Quake-style" valves in microfluidic systems.[44-45] The key is that our approach similarly allows a single, individually-addressable flow channel to remain open within the 3D structure of the device and so operation of our devices is functionally analogous and crucially, scales to similarly to potentially control very large numbers of channels with significantly fewer control inputs. We anticipate that the multilayer 3D-printed valve systems we present can help in the development of microfluidic systems for high-throughput applications in biotechnology and medicine. Further optimization of resins and auxetic cell architectures promises to enable larger, more complex devices and active operations that exploit the unusual properties engineered mechanical metamaterials.

## MATERIAL AND METHODS

### Materials

3D printing was performed using a commercially available Elegoo Mars 2 or Mars 3 printer. The resin consisted of a mixture SuperFlex resin (3D Materials) and Plant-Based UV Resin (SUNLU) in the ratio of 10:1.

### Valve design, fabrication, post-processing, and sterilization

The valves were designed using the OpenSCAD[46-47] software and exported as STL files. The STL files were then sliced using Chitubox (CBD-Tech) with the settings: . Following the printing process, the structures contained unpolymerized resin in the flow channels, which was purged through multiple rinses with isopropyl alcohol (IPA). Subsequently, the printed structures underwent a cleaning and drying process using dry air, followed by sonication in deionized water for 5 minutes. After sonication, the structures were once again dried and cured with a UV lamp for 5 minutes.

### Uniaxial compression test of auxetic valves at different values of θ and φ

Uniaxial compression tests of the specimens at different values of θ and φ were performed using the Torbal force gauge at the crosshead speed of 10mm/min. The structures were compressed corresponding to the distance the inner layers were fully compressed.

### Flow rate measurement in the flow channels

Flow rates within the flow channels in the 3D auxetic valve device used a pressure controller (OB1 Base MkIII, Elveflow, Paris, France connected to a liquid reservoir and BFS1 Coriolis based flow sensor.  The pressure controller, as above exerts a constant pressure to the liquid reservoir which pushes liquid into the auxetic valve. The input to the flow channels in the auxetic valve is connected to the BFS1 flow sensor which measures the fluid flow rate at the given regulator pressure. The reading were performed done for the two configuration of flow channels in the auxetic structures, one "closeable" and one "always open".

## Finite element analysis (FEA)

Models of the auxetic grids were built in the FreeCAD (https://github.com/FreeCAD/FreeCAD) Sketcher workbench with the parameters as shown in Figure 4. As described in the main text, the lengths of the side and re-entrant segments of each auxetic unit was calculated so that 5 unit cells would fit into across is 9 mm x 9 mm total area, with a thickness of 5 mm. A mesh was generated using Netgen. The structure was assigned a Young's modulus of 6.0 GPa and for compressive force of 100 to 1000 N applied to the top of the solved using CalculiX. A Python script was written to generate these structures to perform these calculations values for each θ and φ in increments of 100 N, and then determine for the Y displacement minimum and output the results. Spring constants were then estimated by fitting a line to the linear (low force) segments of the simulated force vs. displacement curves. For the individual auxetic cells (bowties) calculations were performed similarly except simulations were performed only for forces between 1 to 10 N.

## DATA AVAILABILITY

The data that support the findings of this study are available from the corresponding author upon reasonable request. OpenSCAD scripts, STL files for valve designs, and FEA scripts in Python for analysis of these structures will be made available upon publication of this manuscript.

## SUPPLEMENTARY DATA

Additional images of auxetic structurers subjected to applied pressures and force vs. displacement graphs are available in the supplementary data.

## AUTHOR CONTRIBUTIONS

Conceptualization: TS, SM, RT, JRA, EAJ; Data curation: TS, DS, SM, RT; Formal Analysis: TS, DS, SM, RT; Funding acquisition: EAJ; Investigation: TS, DS, SM, RT; Methodology: TS, SM, RT; Project administration: EAJ; Resources EAJ; Software: TS, DS, RT; Supervision: JRA, EAJ; Validation: TS, SM; Visualization TS, SM, RT, EAJ; Writing – original draft: TS, EAJ; Writing – review & editing: TS, DS, SM, RT, JRA, EAJ.

## ACKNOWLEDGEMENTS

We wish to acknowledge Alden Contreras, Rutujaa Kulkarni, and Evan McDowell for helpful technical assistance.

## FUNDING

This research was supported by the National Institute of General Medical Sciences of the National Institutes of Health (1R35GM133483 to EAJ), National Science Foundation award #32027738 (to EAJ) which provided support for TS, and a UNCG internal research grant award (to EAJ). RT was supported by The Draelos Science Scholars Program and SM received support from UNCG Office of the Provost, Division of Student Success. This work was performed in part at the Joint School of Nanoscience and Nanoengineering, a member of the Southeastern Nanotechnology Infrastructure Corridor (SENIC) and National Nanotechnology Coordinated Infrastructure (NNCI), which is supported by the National Science Foundation (Grant ECCS-1542174)

## CONFLICT OF INTEREST

Authors declare no conflicts of interest.